\documentclass{appolb}
\usepackage{graphicx}
\usepackage[english]{babel}
\usepackage[utf8]{inputenc}
\usepackage{graphicx,xcolor}
\usepackage{epsf}
\usepackage{amssymb}
\usepackage{hyperref}
\usepackage{booktabs}
\usepackage{siunitx}
\usepackage{hyperref}
\usepackage{placeins}
\usepackage{epstopdf}
\usepackage{listings} 
\newcommand{\beq}{\begin{equation}}
\newcommand{\eeq}{\end{equation}}
\newcommand{\bea}{\begin{eqnarray}}
\newcommand{\eea}{\end{eqnarray}}

\begin{document}
\title{Mueller-Navelet jets at LHC: 
       discriminating BFKL from DGLAP by asymmetric cuts %
\thanks{Presented at \bf{\textit{EDS Blois 2015: 
        The 16th conference on Elastic and Diffractive scattering}}.} %
}
\author{\underline{F.G.~Celiberto}$^{1}$, D.Yu.~Ivanov$^{2,3}$, 
        B.~Murdaca$^{1}$, A.~Papa$^{1}$
\address{\centerline{${}^1$ {\sl Dipartimento di Fisica, Universit\`a della Calabria,}}
\centerline{\sl and Istituto Nazionale di Fisica Nucleare, Gruppo collegato di
Cosenza,}
\centerline{\sl Arcavacata di Rende, I-87036 Cosenza, Italy}
\vspace{0.1cm}
\centerline{${}^2$ {\sl Sobolev Institute of Mathematics,
630090 Novosibirsk, Russia}}
\vspace{0.1cm}
\centerline{${}^3$ {\sl Novosibirsk State University, 630090 Novosibirsk, Russia}}
}}
\maketitle
\begin{abstract}
The Mueller-Navelet di-jet production process 
represents an ultimate testfield of pQCD in the
high-energy limit. Several experimental analyses carried out so far 
are in good agreement with theoretical predictions,
based on collinear factorization and BFKL resummation
of energy logarithms in the next-to-leading approximation,
with the CMS experimental data at center-of-mass energy equal to 7 TeV. 
However, the question if the same data can
be described also by fixed-order perturbative approaches has
not yet been fully answered. We discuss 
how the use of partially asymmetric cuts in the transverse momenta 
of the detected jets allows to discriminate 
between BFKL-resummed and fixed-order predictions 
(the latter in the high-energy limit) 
in observables related with this process at LHC.
\end{abstract}
  

\section{Introduction}
\label{intro}

The inclusive hadroproduction of two jets 
featuring transverse momenta of the same order and much larger than the 
typical hadronic masses and being separated by a large rapidity gap $Y$, 
the so-called Mueller-Navelet jets~\cite{Mueller:1986ey}, is a fundamental 
testfield for perturbative QCD in the high-energy limit.
At the LHC energies, the theoretical description of this process lies
between two distinct approaches: collinear factorization 
and BFKL~\cite{BFKL} resummation. On one side, at leading twist the
process can be seen as the hard scattering of two partons, each emitted by one
of the colliding hadrons according to the appropriate parton distribution 
function (PDF), see Fig.~1 of~\cite{Celiberto_jets:2015}. 
Collinear factorization takes care
to resum the logarithms of the hard scale, through the 
standard DGLAP evolution~\cite{DGLAP} of the PDFs and the fixed-order 
radiative corrections to the parton scattering cross section.
The other approach is the BFKL resummation of energy 
logarithms, which are so large to compensate the small QCD coupling and must 
therefore be accounted for to all orders. 
These logarithms are related with the emission of undetected partons between the two 
jets (the larger $s$, the larger the number of partons), which lead to
a reduced azimuthal correlation between the two detected forward jets,
in comparison to the fixed-order DGLAP calculation, where jets are emitted
almost back-to-back.
In the BFKL approach energy logarithms are systematically resummed in the
leading logarithmic approximation (LLA) and in the next-to-leading logarithmic approximation 
(NLA).
To get the cross section, the BFKL Green's function must be convoluted with
two impact factors for the transition from the colliding parton to the
forward jet. 
They were first calculated 
with NLO accuracy in~\cite{Bartels:2002yj} and the result was later 
confirmed in~\cite{Caporale:2011cc}. A simpler expression, more 
practical for numerical purposes, was obtained in~\cite{Ivanov:2012ms}
adopting the so-called ``small-cone'' approximation 
(SCA)~\cite{Furman:1981kf,Aversa}.
Unfortunately, the NLO BFKL corrections
for the $n=0$ conformal spin are with opposite sign with respect to the
leading order (LO) result and large in absolute value~\cite{Ivanov2006}.
This calls for some optimization procedure.
Common optimization methods are those inspired by the \emph{principle of minimum 
sensitivity} (PMS)~\cite{PMS}, the \emph{fast apparent convergence} 
(FAC)~\cite{FAC} and the \emph{Brodsky-LePage-Mackenzie method} 
(BLM)~\cite{BLM}.
This variety of options reflects in the large number of numerical studies
of the Mueller-Navelet jet production process at LHC, both at a center-of-mass
energy of 14~TeV~\cite{Colferai2010,Caporale2013,Salas2013} and  
7~TeV~\cite{Ducloue2013,Ducloue2014,Ducloue:2014koa,Caporale:2014gpa}.
In the case of \emph{asymmetric} cuts, the Born term, 
present only for back-to-back jets, is suppressed and the effects of the additional undetected 
hard gluon radiation is enhanced, thus making more visible the BFKL
resummation, with respect to  DGLAP calculations, 
in all observables involving $C_0$~\cite{Caporale:2014gpa}.
So, we compare predictions for several azimuthal correlations 
and their ratios obtained, on one side, by a fixed-order high-energy
DGLAP calculation at the NLO and, on the other side, 
by BFKL resummation in the NLA. 
To single out the only effect of transverse momentum cuts, 
we consider just one optimization procedure 
(the BLM one, in the two variants discussed in~\cite{BLMpaper}).

\section{Theoretical setup}
\label{theory}

The process under exam is the production of Mueller-Navelet 
jets~\cite{Mueller:1986ey} in proton-proton collisions
\begin{eqnarray}
\label{process}
p(p_1) + p(p_2) \to {\rm jet}(k_{J_1}) + {\rm jet}(k_{J_2})+ X \;,
\end{eqnarray}
where the two jets are characterized by high transverse momenta,
$\vec k_{J_1}^2\sim \vec k_{J_2}^2\gg \Lambda_{QCD}^2$ and large separation
in rapidity, while $p_1$ and $p_2$ are taken as Sudakov vectors.
The cross section of the process can be presented as
\beq
\frac{d\sigma}
{dy_{J_1}dy_{J_2}\, d|\vec k_{J_1}| \, d|\vec k_{J_2}|
d\phi_{J_1} d\phi_{J_2}}
=\frac{1}{(2\pi)^2}\left[{\cal C}_0+\sum_{n=1}^\infty  2\cos (n\phi )\,
{\cal C}_n\right]\, ,
\eeq
where $\phi=\phi_{J_1}-\phi_{J_2}-\pi$, while ${\cal C}_0$ gives the total
cross section and the other coefficients ${\cal C}_n$ determine 
the distribution of the azimuthal angle of the two jets. 
We concentrate on 
the so-called {\it exponentiated} representation, and use
the BLM optimization procedure, {\it i.e.} we choose
the scale $\mu_R$ such that it makes vanish completely 
the $\beta_0$-dependence of a given observable. 
As discussed in~\cite{Caporale:2014gpa}, we implement
two variants of the BLM method, dubbed $(a)$ and $(b)$~\cite{BLMpaper}. 
A common optimal value for the renormalization scale $\mu_R$ 
and for the factorization scale $\mu_F$ is used.
In~\cite{Caporale:2014gpa} it was shown that this setup 
allows a nice agreement with CMS data for several azimuthal correlations and 
their ratios in the large $Y$ regime.
The BFKL and DGLAP expressions for the coefficients ${\cal C}_n$,
in the two variants of BLM setting, 
are given in Eqs.~(4), (6), (12) and~(13) of Ref.~\cite{Celiberto_jets:2015}.
We note that, in our way to implement the BLM procedure (see~\cite{BLMpaper}),
the final expressions are given in terms of $\alpha_s$ in the 
$\overline{\rm MS}$ scheme, although in one intermediate step the MOM scheme
was used.

\section{Numerical analysis}
\label{results}

\begin{figure}[p]
\centering
   \includegraphics[scale=0.35]{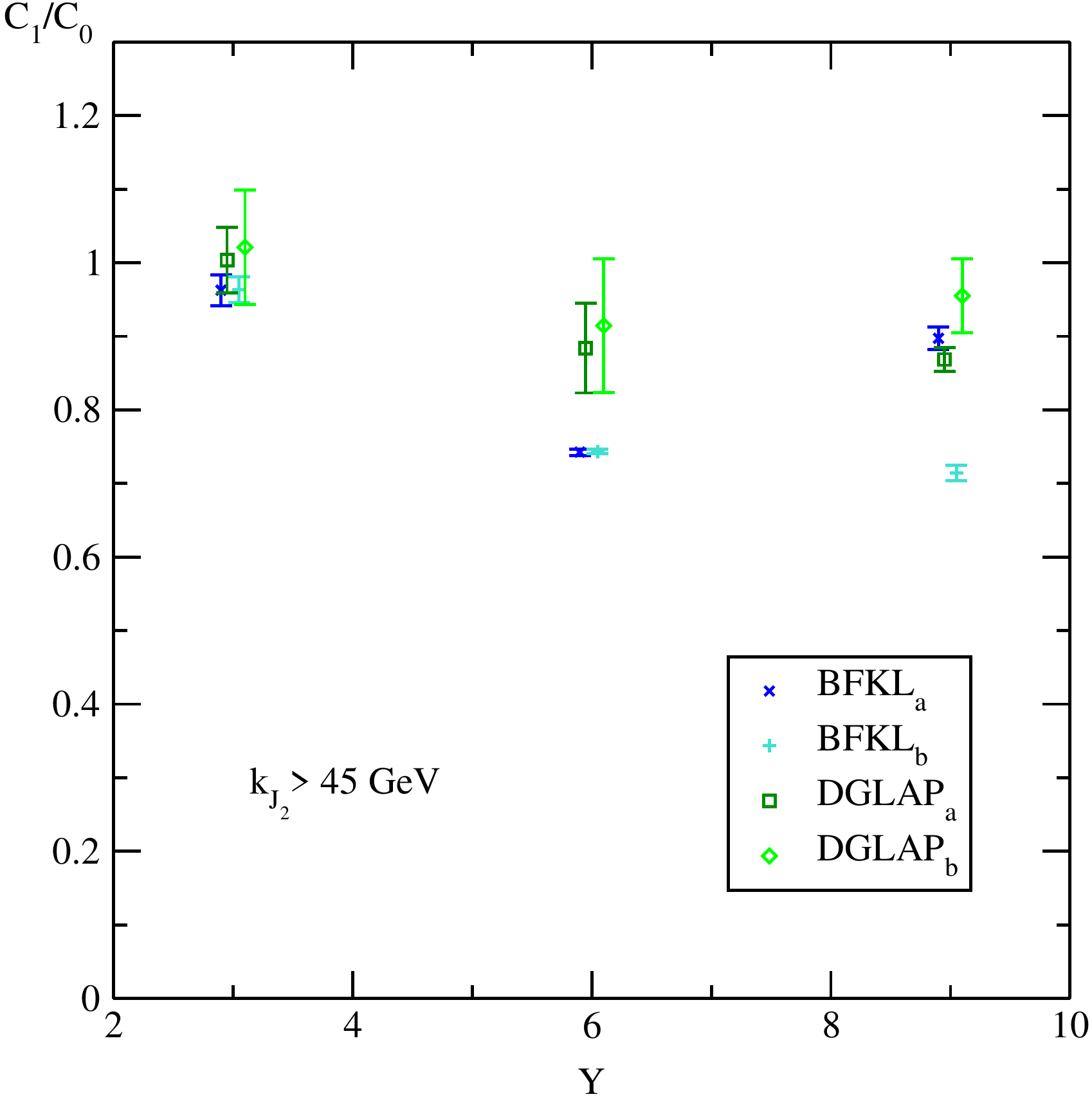}
   \includegraphics[scale=0.35]{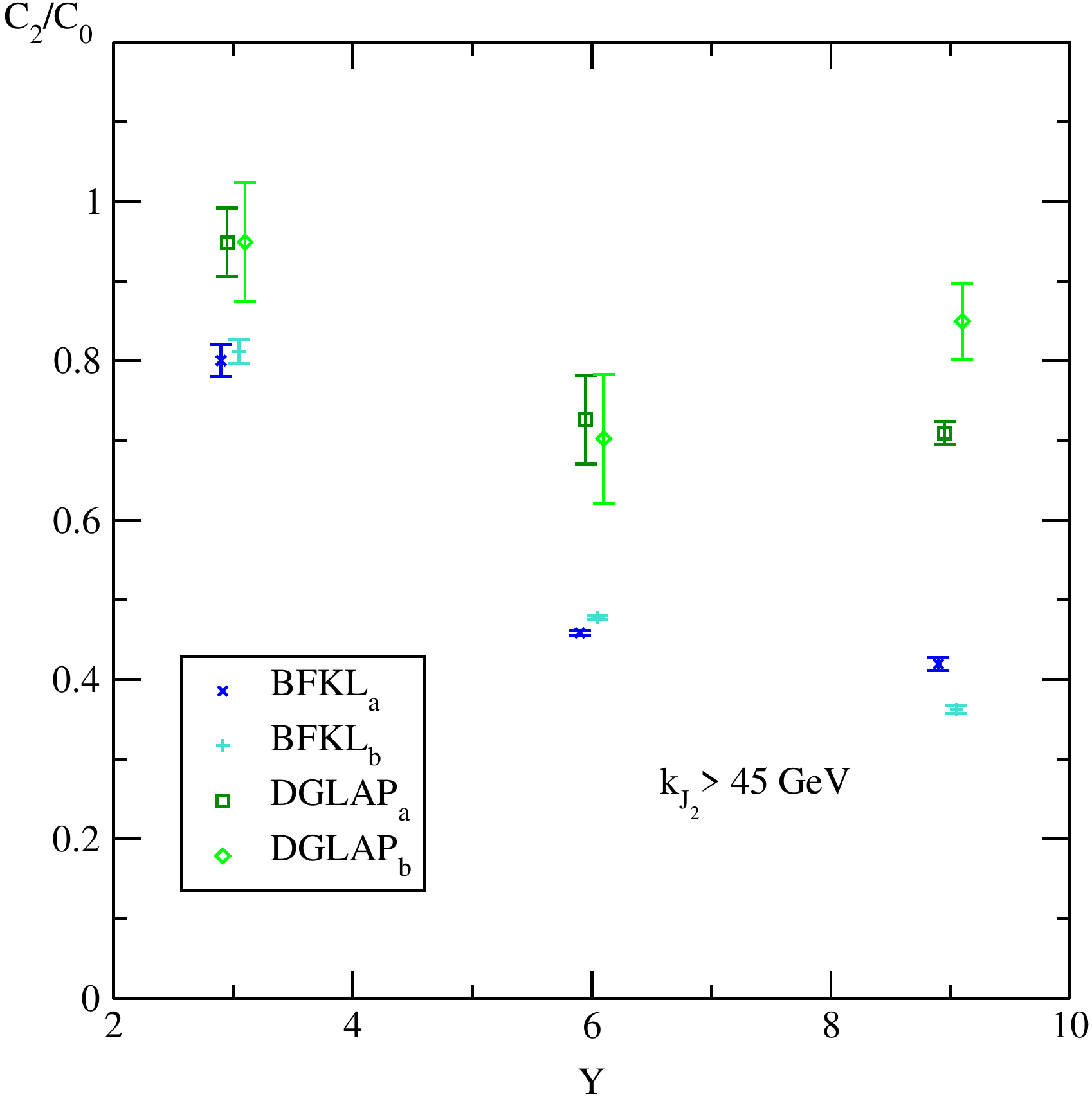}

   \includegraphics[scale=0.35]{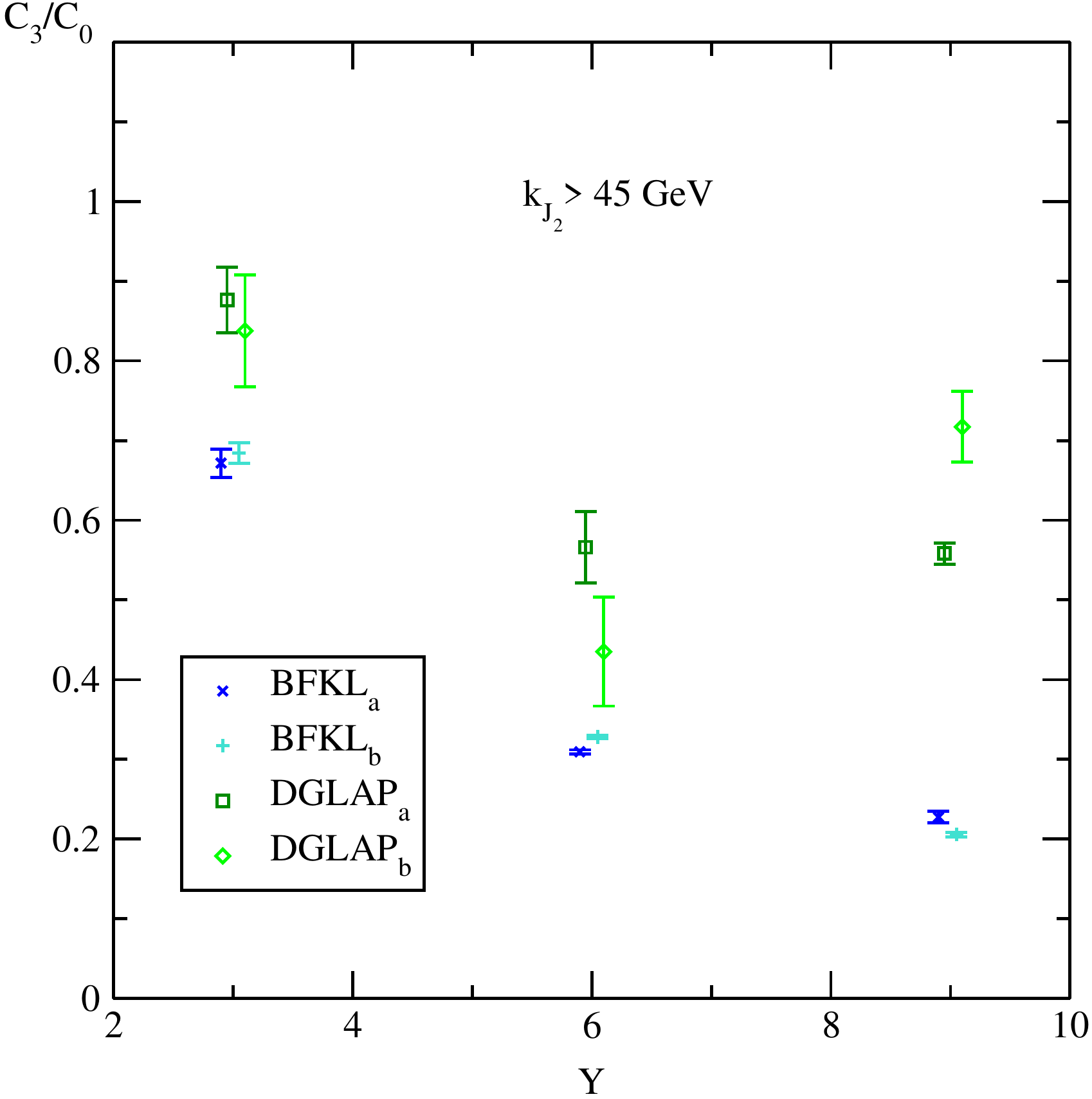}
   \includegraphics[scale=0.35]{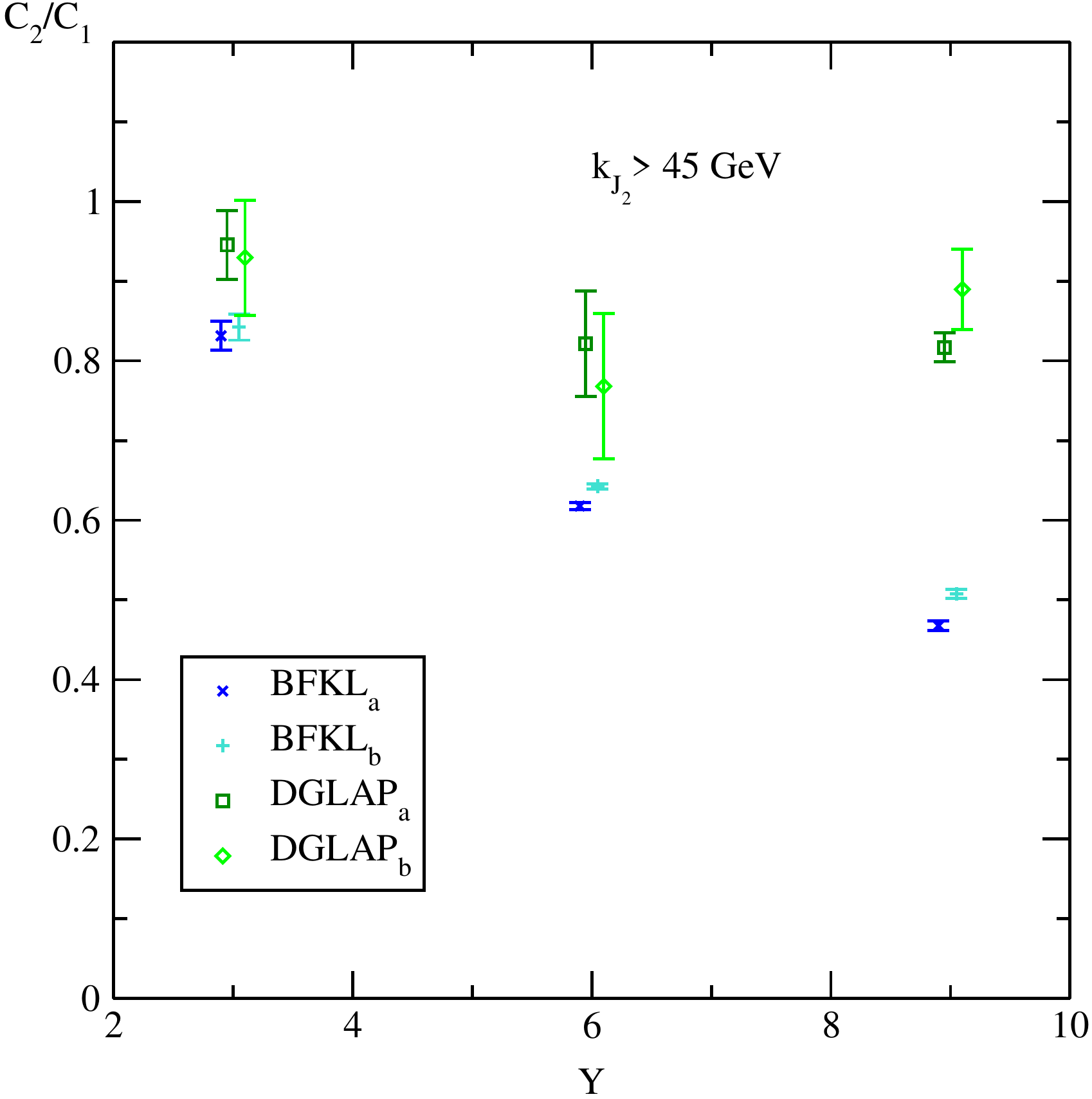}

   \includegraphics[scale=0.35]{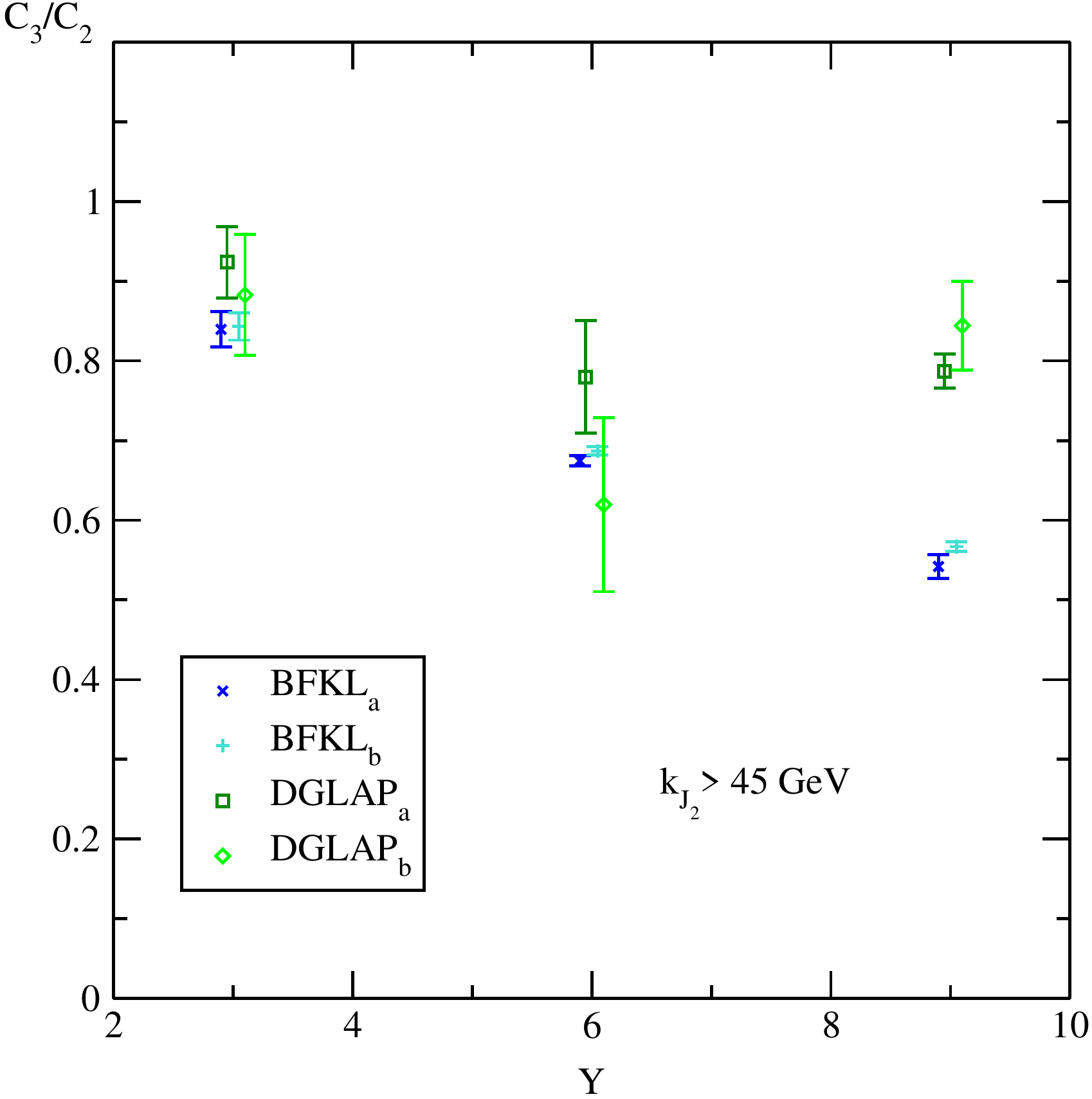}
\caption{\small $Y$-dependence of several ratios $C_m/C_n$ for $k_{J_1,\rm min}=35$~GeV and $k_{J_2,\rm min}=45$~GeV, for BFKL and DGLAP in the two variants of the 
BLM method (data points have been slightly shifted along the horizontal
axis for the sake of readability). 
For the numerical values, see Table 1 of Ref.~\cite{Celiberto_jets:2015}.}
\label{plot45}
\end{figure}

We present our results for the dependence on the
rapidity separation between the detected jets, $Y=y_{J_1}-y_{J_2}$, 
of ratios ${\cal R}_{nm}\equiv{\cal C}_n/{\cal C}_m$ between the 
coefficients ${\cal C}_n$. Among them, the ratios of
the form $R_{n0}$ have a simple physical interpretation, being the azimuthal
correlations $\langle \cos(n\phi)\rangle$.
In order to match the kinematic cuts used by the CMS collaboration, we will
consider the \emph{integrated coefficients} 
given in Eq.~(13) of Ref.~\cite{Caporale:2014gpa}
and their ratios $R_{nm}\equiv C_n/C_m$. We will take jet rapidities in the
range delimited by $y_{1,\rm min}=y_{2,\rm min}=-4.7$  and 
$y_{1,\rm max}=y_{2,\rm max}=4.7$ 
and consider $Y=3$, 6 and 9. 
As for the jet transverse momenta we make two \emph{asymmetric} choices: 
(1) $k_{J_1,\rm min} = 35$~GeV, $k_{J_2,\rm min} = 45$~GeV 
(Fig.~\ref{plot45}) and 
(2) $k_{J_1,\rm min} = 35$~GeV, $k_{J_2,\rm min} = 50$~GeV
(Fig.~3 of Ref.~\cite{Celiberto_jets:2015}).
The center-of-mass energy is fixed at $\sqrt s=7$~TeV. 
We can clearly see that, at 
$Y=9$, BFKL and DGLAP, in both variants $(a)$ and $(b)$ of the
BLM setting, give quite different predictions for the all considered ratios 
except $C_1/C_0$; at $Y=6$ this happens in fewer cases, while at $Y=3$ 
BFKL and DGLAP cannot be distinguished with the given uncertainties. This scenario
is similar in the two choices of the transverse momentum cuts.
For a detailed discussion of the numerical tools used and 
of the uncertainty estimation in our analysis, 
see sections~3.2 and~3.3 of Ref.~\cite{Celiberto_jets:2015}.



\section{Conclusions}
\label{conclusions}

In this paper we considered the Mueller-Navelet jet production
process at LHC at the center-of-mass energy of 7~TeV and compared
predictions for several azimuthal correlations and ratios between them,
both in full NLA BFKL approach and in fixed-order NLO DGLAP.
Differently from current experimental analyses of the same process, 
we have used {\it asymmetric} cuts for the transverse
momenta of the detected jets. 
The use of {\it symmetric} cuts for jet
transverse momenta maximizes the contribution of the Born term, which is present
for back-to-back jets only and is expected to be large, therefore making
less visible the effect of the BFKL resummation. This phenomenon
could be at the origin of the instabilities observed in the NLO fixed-order
calculations of~\cite{Andersen:2001kta,Fontannaz:2001nq}.
Another important benefit from the use of asymmetric cuts, pointed out 
in~\cite{Ducloue:2014koa}, is that the effect of violation of the 
energy-momentum conservation in the NLA is strongly suppressed with respect 
to what happens in the LLA.
In view of all these considerations, we strongly suggest experimental 
collaborations to consider also asymmetric cuts in jet transverse momenta in all
future analyses of Mueller-Navelet jet production process.

\vspace{1.0cm} \noindent
{ \bf Acknowledgments} \vspace{0.5cm}

\small

\indent 
The work of D.I. was also supported in part 
by the grant RFBR-15-02-05868-a. 
\indent
The work of B.M. was supported in part by the grant RFBR-13-02-90907 and by 
the European Commission, European Social
Fund and Calabria Region, that disclaim any liability for the use 
that can be done of the information provided in this paper.

\normalsize



\end{document}